%% file: ANA-HIGG-2018-13-PAPER.tex
\pdfoutput=1
\newcommand*{\ATLASLATEXPATH}{}
\documentclass[cernpreprint, texlive=2016, UKenglish]{\ATLASLATEXPATH atlasdoc}
 
\usepackage[csquotes=false,default,subfigure]{\ATLASLATEXPATH atlaspackage}
\usepackage{rotating}
\usepackage{pdflscape}
\usepackage{\ATLASLATEXPATH atlasbiblatex}
 
\usepackage{\ATLASLATEXPATH atlasphysics}
 
\graphicspath{{logos/}{figures/}}
 
\usepackage{ANA-HIGG-2018-13-PAPER-defs}

\AtlasTitle{Observation of Higgs boson production in association with a top quark pair at the LHC with the ATLAS detector}

\AtlasAbstract{
The observation of Higgs boson production in association with a top quark pair (\ttH), based on the analysis of
proton--proton collision data at a centre-of-mass energy of 13~\tev\ recorded with the ATLAS detector at the Large Hadron
Collider, is presented. Using data corresponding to integrated luminosities of up to 79.8~\ifb, and considering Higgs boson decays into \bbbar, \WW, \tautau, \yy, and \ZZ,
the observed significance is \SignCombRunTwoObs\ standard deviations, compared to an expectation of \SignCombRunTwoExp\ standard deviations.
Combined with the \ttH\ searches using a dataset corresponding to integrated luminosities of 4.5~\ifb\ at 7~\tev\ and 20.3~\ifb\ at 8~\tev, the observed (expected) significance is \SignCombRunOneRunTwoObs\ (\SignCombRunOneRunTwoExp)  standard deviations.
Assuming Standard Model branching fractions, the total \ttH\ production cross section at 13~\tev\ is measured to be \XSComb~fb, in agreement with the Standard Model prediction.
}
 
\author{The ATLAS Collaboration}
 
\AtlasRefCode{HIGG-2018-13}
 
\PreprintIdNumber{CERN-EP-2018-138}

\AtlasDate{\today}

\AtlasJournalRef{Phys. Lett. B 784 (2018) 173}
\AtlasDOI{10.1016/j.physletb.2018.07.035}

\hypersetup{pdftitle={ATLAS document},pdfauthor={The ATLAS Collaboration}}
 
\begin{document}
 
\maketitle
 
\section{Introduction}
% The next lines are included from the .//sections/introduction.tex input file
After the discovery of the Higgs boson in 2012 by the ATLAS and CMS Collaborations~\cite{HIGG-2012-27,CMS-HIG-12-028}, many measurements of its properties were performed~\cite{HIGG-2015-07,HIGG-2016-29,Aaboud:2017vzb,HIGG-2016-21,Sirunyan:2017khh,Sirunyan:2017elk}. No significant deviations from the Standard Model (SM) predictions were found.
A probe of fundamental interest to further explore the nature of the Higgs boson is its coupling to the top quark, the heaviest particle in the SM. Indirect measurements of the Yukawa coupling between the Higgs boson and the top quark were made by the ATLAS and CMS Collaborations~\cite{HIGG-2015-07}, assuming no contribution from unknown particles in the gluon--gluon fusion (ggF) loop. A more direct test of this coupling can be performed through the production of the Higgs boson in association with a top quark pair, \ttH. Using  a proton--proton ($pp$) dataset corresponding to an integrated luminosity of 36.1~$\pm$~0.8~\ifb~\cite{DAPR-2013-01}, at a centre-of-mass energy $\sqrt{s}=13$~\tev, evidence of this production mode was found in 2017 by the ATLAS Collaboration~\cite{HIGG-2017-02}, with an observed (expected) significance relative to the background-only hypothesis of 4.2 (3.8) standard deviations. Combining data at 7, 8, and 13~\tev, the CMS Collaboration reported an observed (expected) significance
of 5.2 (4.2) standard deviations~\cite{CMS-tth}.
 
This Letter presents results of the search for the \ttH\ process and the measurement of the \ttH\ production cross section using data produced in $pp$ collisions by the Large Hadron Collider (LHC) and recorded with the ATLAS detector. The ATLAS detector is described in detail in Refs.~\cite{PERF-2007-01,ATLAS-TDR-19}.
Compared to Ref.~\cite{HIGG-2017-02}, the $H\to \yy$ and $H\to \ZZfourl$ ($\ell = e, \mu$) analyses are updated with the 13~\tev\
data collected in 2017.
Improved lepton and photon reconstruction algorithms~\cite{ATL-PHYS-PUB-2017-022} and analysis techniques are used.
The updated analyses are combined with the $H\to \bbbar$ and multilepton analyses from Refs.~\cite{HIGG-2017-03,HIGG-2017-02}, the latter targeting Higgs boson decays into \WW, $H\to\tautau$ with hadronically and leptonically decaying $\tau$-leptons, and $H\to \ZZ$ without \ZZfourl.
Furthermore, a combination is performed with the results based on 4.5~$\pm$~0.4~\ifb\ and 20.3~$\pm$~0.1~\ifb\ of $pp$ data recorded in 2011 and 2012 at $\sqrt{s}=7$~\tev\ and $\sqrt{s}=8$~\tev\ respectively~\cite{HIGG-2014-06,HIGG-2015-05,HIGG-2013-27,HIGG-2013-26,HIGG-2013-25}. A Higgs boson mass corresponding to the measured value of 125.09~$\pm$~0.24~\gev~\cite{HIGG-2014-14} is assumed everywhere.

% End of text imported from the .//sections/introduction.tex input file
\section{\hyy}
% The next lines are included from the .//sections/gammagamma.tex input file
\newcommand{\AtlasCoordFootnote}{
ATLAS uses a right-handed coordinate system with its origin at the nominal interaction point (IP) in the centre of the detector and the $z$-axis along the beam pipe.  The $x$-axis points from the IP to the centre of the LHC ring, and the $y$-axis points upwards.  Cylindrical coordinates $(r,\phi)$ are used in the transverse plane, $\phi$ being the azimuthal angle around the $z$-axis.  The pseudorapidity is defined in terms of the polar angle $\theta$ as $\eta = -\ln \tan(\theta/2)$.  Angular distance is measured in units of $\Delta R \equiv \sqrt{(\Delta\eta)^{2} + (\Delta\phi)^{2}}$.}
 
In the $H\to\gamma\gamma$ analysis, using a dataset corresponding to an integrated luminosity of 79.8~$\pm$~1.6~\ifb\ at $\sqrt{s}=13$~\tev, events with two isolated photon candidates with transverse momenta\footnote{\AtlasCoordFootnote} \pt\ larger than 35~\GeV\ and 25~\GeV\ are selected. Both photons must satisfy the quality requirements discussed in Ref.~\cite{HIGG-2016-21}; the diphoton \myy\ invariant mass must be in the range $\myy \in [105-160]\,\GeV$, and the leading (subleading) photon must have $\pt/\myy$ > 0.35 (0.25).
At least one jet with $\pt > 25$~\GeV\ and containing a $b$-hadron, identified using a $b$-tagging algorithm with an efficiency of 77\%~\cite{PERF-2012-04,ATL-PHYS-PUB-2015-022,Aaboud:2018xwy}, is required.
Two signal regions targeting $t\bar{t}H$ production are defined. One is enriched in hadronic top-quark decays by requiring at least two additional jets and zero isolated leptons (electrons or muons). This `\had' region contains events where both top quarks decay into hadrons or the leptons from decays of the top quarks are not reconstructed or identified. The `\lep' region is instead enriched in semileptonic top-quark decays by requiring events to have at least one isolated lepton.

The sensitivity of the analysis is improved relative to Ref.~\cite{HIGG-2016-21}. Two dedicated boosted decision trees (BDTs) are trained using the \textsc{XGBoost} package \cite{XGBoost} to discriminate the $t\bar{t}H$ signal from the main background processes. These are non-resonant diphoton production processes, including  \ttbar\ production together with a photon pair. The background processes also include non-\ttH\ Higgs boson production: mainly associated production with a single top quark $tH$ and ggF in the \had\ region, and $tH$ and associated production with a vector boson $VH$, where $V = W, Z$, in the \lep\ region. The \ttH, ggF, vector-boson fusion (VBF), and $VH$ production processes were simulated with \textsc{Powheg}+\textsc{Pythia8}~\cite{powheg1,powheg2,powheg3,powheg5,Campbell:2012am,Luisoni:2013kna,Hamilton:2013fea,pythia,pythia8}. The production of a Higgs boson in association with two $b$-quarks, \bbH, and $tH$ were modelled using \textsc{Madgraph5\_aMC@NLO}+\textsc{Pythia8}~\cite{Alwall:2014hca,Wiesemann:2014ioa}.
The BDT in the \lep\ region is trained with simulated \ttH\ events, and with background events from a data control region that differs from the \lep\ region by requiring exactly zero $b$-tagged jets, at least one jet, and at least one photon failing either identification or isolation requirements.
This BDT uses the transverse momentum \pt, the pseudorapidity $\eta$, the azimuthal angle $\phi$, and the energy $E$ of up to four (two) leading jets (leptons) in \pt.
It was verified that the BDT is not sensitive to the value of the jet mass. Furthermore, the BDT uses the magnitude and the azimuthal angle $\phi$ of the missing transverse momentum \MET, the transverse momentum of each of the two photons divided by the diphoton invariant mass $\pt$/$m_{\gamma\gamma}$, as well as the $\eta$ and $\phi$ of each photon.
The BDT in the \had\ region is also trained with simulated \ttH\ signal events, and with background events from a data control region with the same selection as the \had\ region, except that at least one photon has to fail either identification or isolation requirements. This BDT uses the \pt, $\eta$, $\phi$, $E$ and the $b$-tagging decision of up to six leading jets, plus the \MET\ information and the same photon observables as used by the BDT in the \lep\ region.
In the \had\ region, the \MET\ information adds discriminating power due to semileptonic top-quark decays with undetected leptons.
The data control regions for the \had\ and \lep\ BDT training are chosen with the goal to maximize the expected sensitivity, which is affected by the number of events in the training sample and background composition.
Events with low values of the BDT response are removed: about 85\% (97\%) of the \ttH\ signal events are selected and about 89\% (43\%) of the non-resonant background events are rejected in the \had\ (\lep) region.
The remaining events are categorized into four (three) bins in the \had\ (\lep) region depending on the value of the BDT response.
The number and boundaries of the BDT bins are chosen to optimize the expected sensitivity to the \ttH\ signal. Figure~\ref{fig:BDT_response_gamgam} shows the distribution of the BDT response for simulated \ttH\ signal, simulated non-\ttH\ Higgs boson production and non-resonant background from data in the diphoton invariant-mass sideband regions $\myy \in [105-120]\,\GeV$ and $\myy \in [130-160]\,\GeV$.
 
\begin{figure}
\centering
\subfigure[ ]{\includegraphics[width=0.48\textwidth]{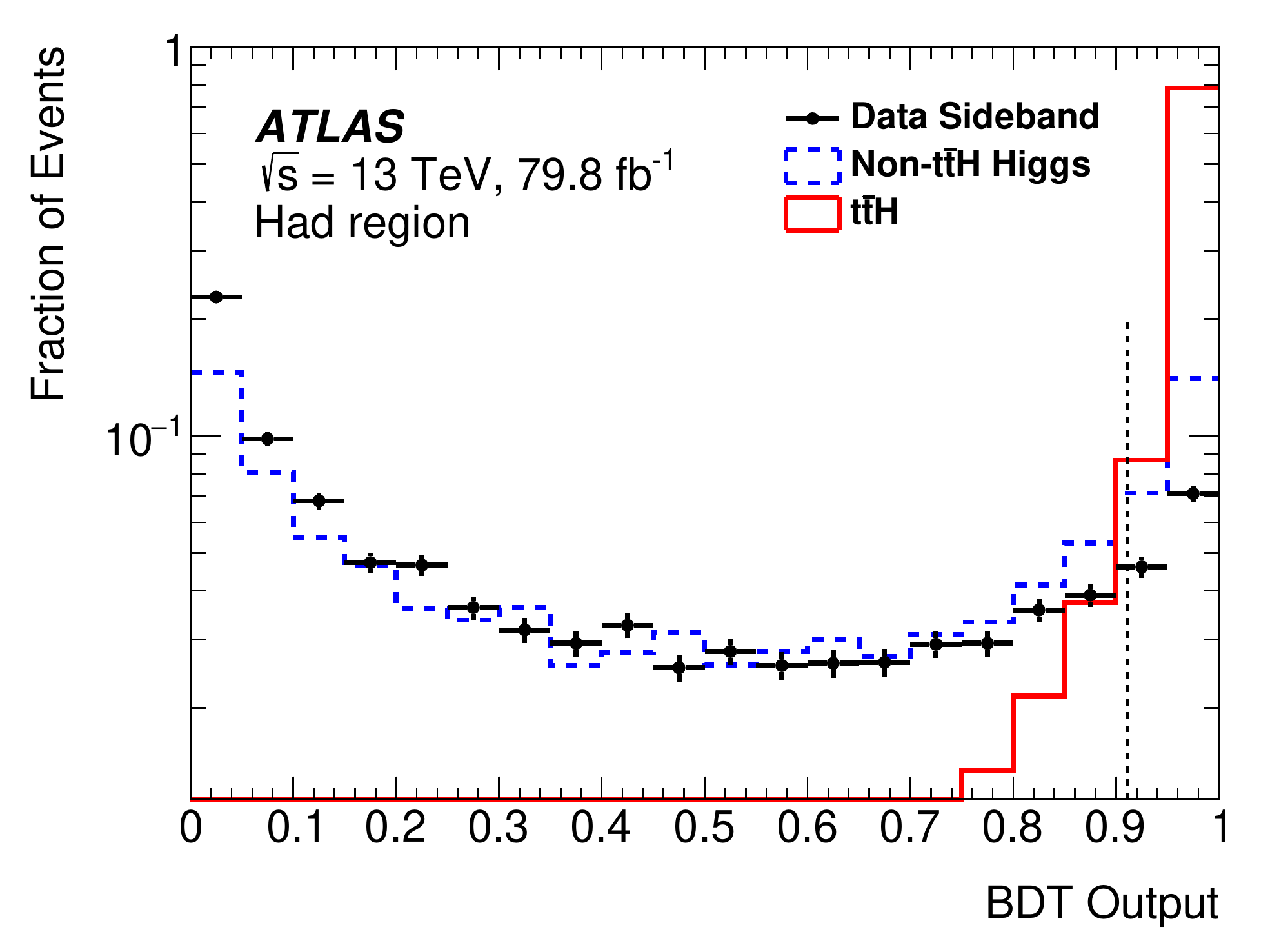}}
\subfigure[ ]{\includegraphics[width=0.48\textwidth]{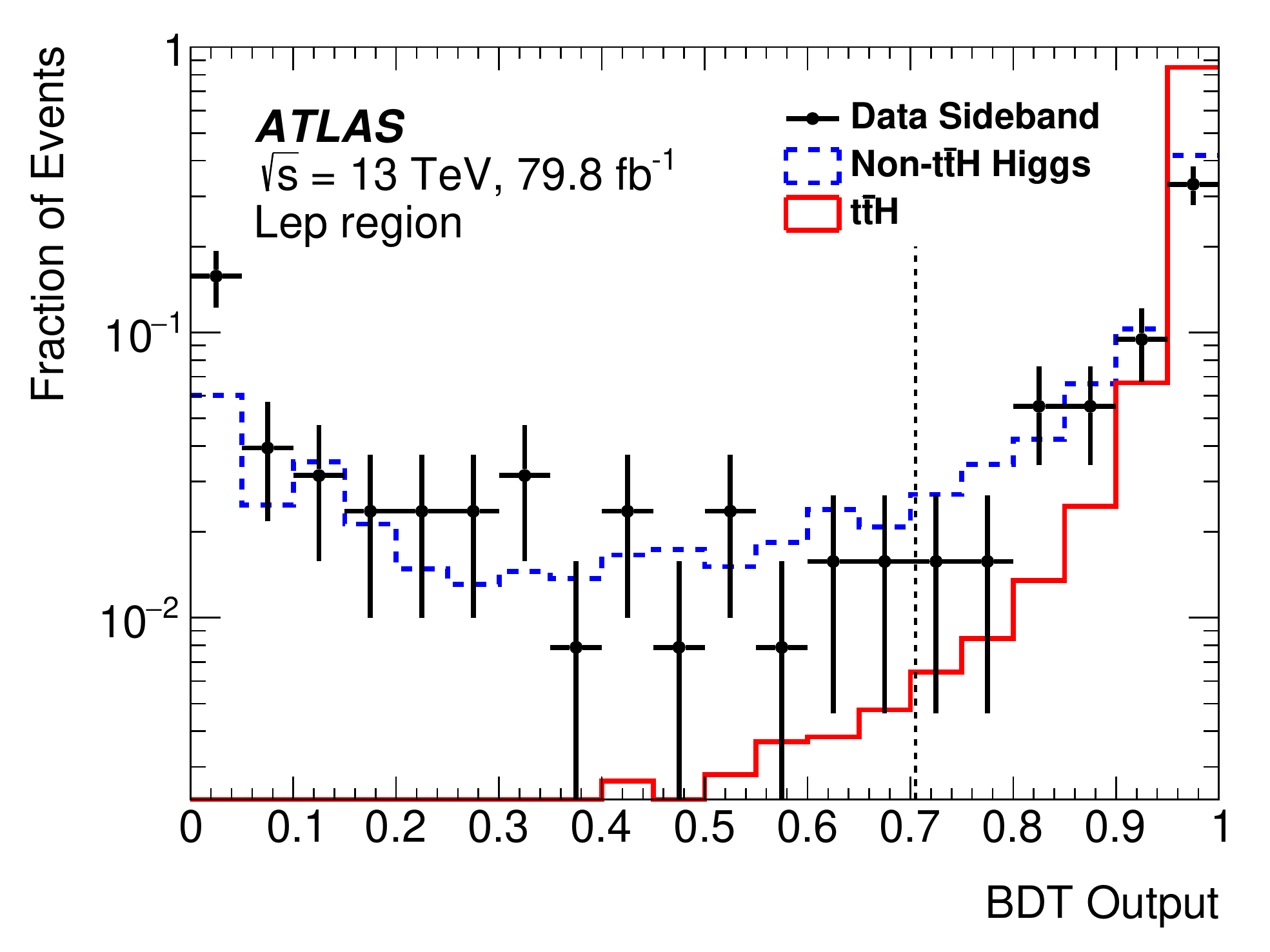}}
\caption{Distribution of the BDT output in the (a) \had\ and (b) \lep\ region in the \hyy\ analysis. The distribution of the simulated \ttH\ signal is compared with that of the other Higgs boson production modes, as well as to the continuum background from data in the diphoton invariant-mass sidebands of $105~\gev < \myy < 120~\gev$ and $130~\gev < \myy < 160~\gev$. Events to the left of the vertical line are rejected. The distributions are normalized to unity.
\label{fig:BDT_response_gamgam}}
\end{figure}

In each BDT bin, the \ttH\ signal yield is measured using a combined unbinned maximum-likelihood fit
to the diphoton invariant mass spectrum in the range 105\,\GeV~$<m_{\gamma\gamma}$~$<$~160\,\GeV, constraining the Higgs boson mass to $125.09 \pm 0.24$~\GeV.
Signal and background shapes are modelled by analytical functions as discussed in Ref.~\cite{HIGG-2016-21}. The functions modelling the Higgs boson signal, used for both the \ttH\ signal and the resonant background from the other Higgs boson production modes, are based on the simulated $m_{\gamma\gamma}$ distributions.  The functional form used to model the continuum background distribution in each BDT bin is chosen using simulated background events for the \lep\ region and a dedicated data control region for the \had\ region, following the procedure described in Refs.~\cite{HIGG-2012-27, HIGG-2016-21}.
This procedure imposes stringent conditions on potential biases in the extracted signal yield, in order to avoid losses in sensitivity.
No evidence of such a bias is observed within the statistical accuracy of the available control samples.
Depending on the BDT bin, either a power-law or an exponential function is chosen, each with one parameter determining the functional shape, and one accounting for the overall background normalization.
The parameters of the continuum background model are left free in the fit. The contributions from the non-\ttH\ production modes are fixed  to their SM expectations~\cite{deFlorian:2016spz,Alwall:2014hca,Wiesemann:2014ioa,powheg1,powheg2,powheg3,powheg5,Campbell:2012am,Luisoni:2013kna,Hamilton:2013fea,pythia,pythia8}. The predicted ggF, VBF and $VH$ (both $qq \rightarrow ZH$ and $gg \rightarrow ZH$) yields are each assigned a conservative 100\% uncertainty, which is due to the theoretical uncertainty in the radiation of additional heavy-flavour jets in these Higgs boson production modes. This is supported by measurements using \hfourl~\cite{HIGG-2016-25}, $t\bar{t}b\bar{b}$~\cite{TOPQ-2014-10}, and $Vb$~\cite{STDM-2012-15,STDM-2012-11} events. The impact of this uncertainty on the \hyy\ and combined results is small.
 
\begin{figure}
\centering
\includegraphics[width=0.8\textwidth]{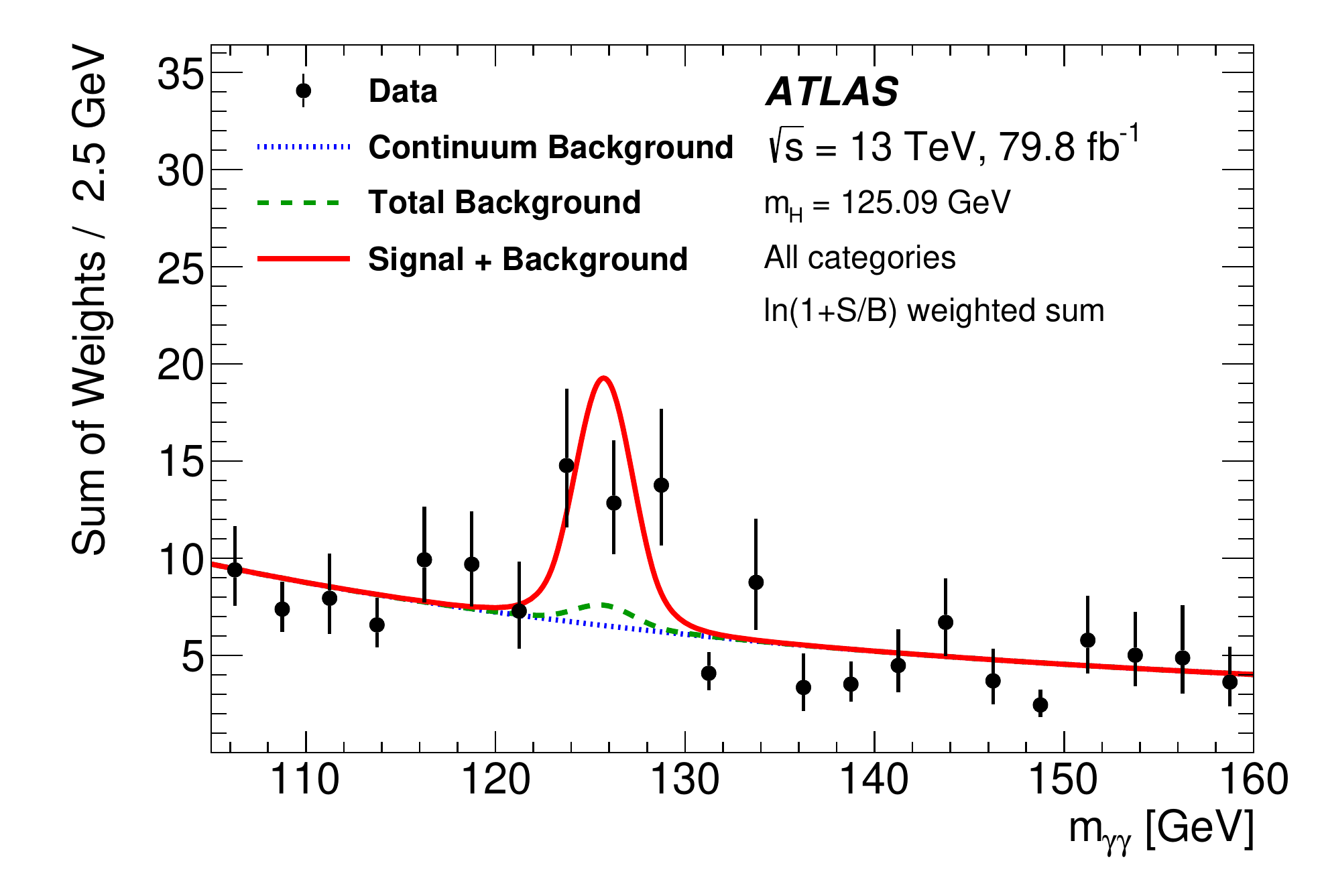}
\caption{
Weighted diphoton invariant mass spectrum in the \ttH-sensitive BDT bins observed in 79.8~\ifb\ of 13~\tev\ data.
Events are weighted by $\ln(1 + S_{90}/B_{90})$, where $S_{90}$ ($B_{90}$) for each BDT bin is the expected \ttH\ signal (background) in the smallest $m_{\gamma\gamma}$ window containing 90\% of the expected signal.
The error bars represent 68\% confidence intervals of the weighted sums. The solid red curve shows the fitted signal-plus-background model with the Higgs boson mass constrained to 125.09 $\pm$ 0.24 \GeV. The non-resonant and total background components of the fit are shown with the dotted blue curve and dashed green curve.
Both the signal-plus-background and background-only curves shown here are obtained from the weighted sum of the individual curves in each BDT bin.
}
\label{fig:myy_ttH_weighted}
\end{figure}

\begin{figure}
\centering
\includegraphics[width=0.8\textwidth]{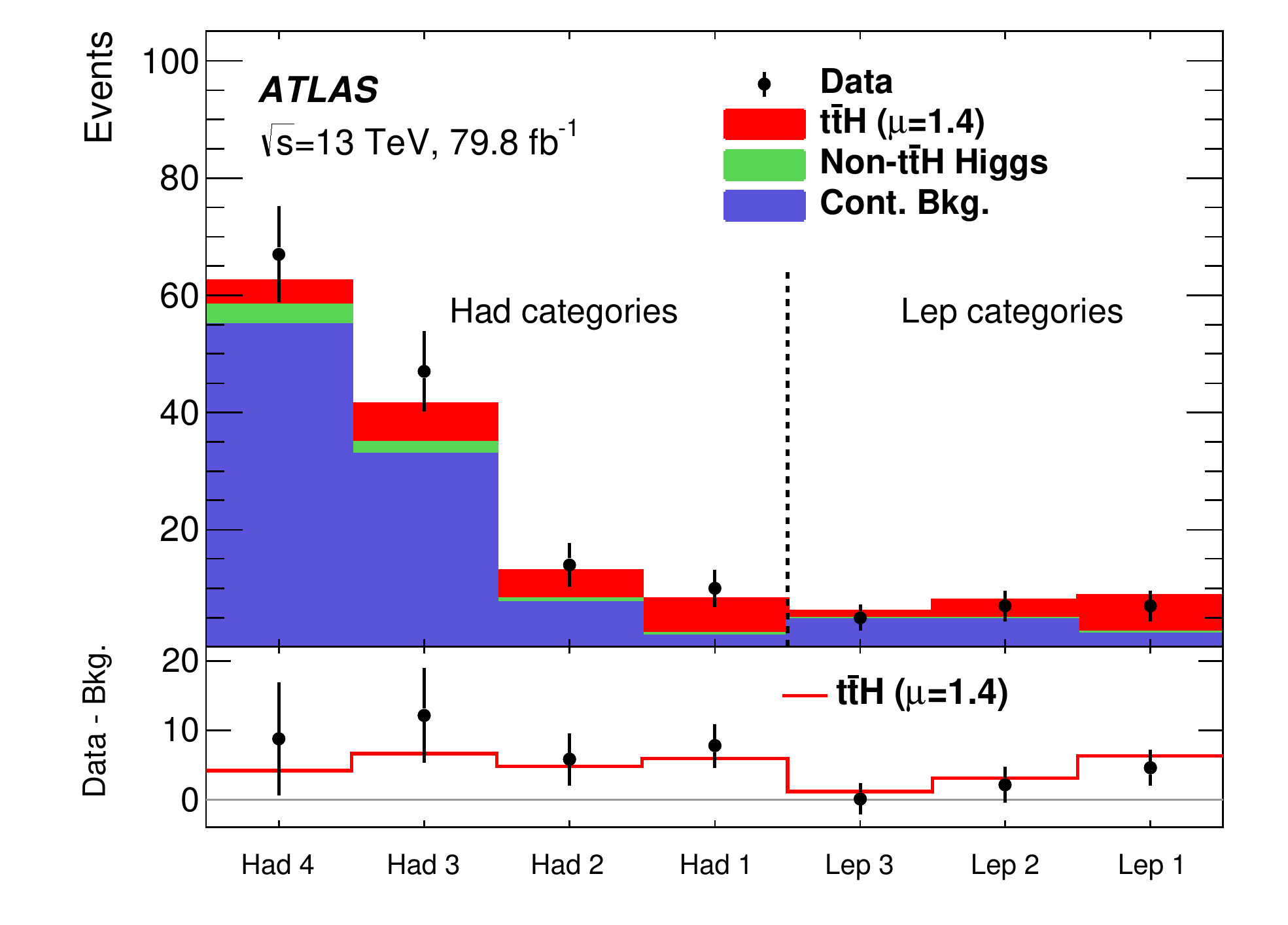}
\caption{ Number of data events in the different BDT bins of the \hyy\ analysis, in the smallest diphoton mass window that contains 90\% of the \ttH\ signal. The expected background and \ttH\ signal (for a signal strength $\mu = \sigma / \sigma_\mathrm{SM}$ of 1.4) are shown as well.  The expected continuum background is extracted from the diphoton mass fits.  The lower panel shows the residuals between the data and the background. The red line shows the expected signal.
The BDT bins are shown in ascending order of signal purity.
\label{fig:stacks}}
\end{figure}

The most important theoretical uncertainties affecting the \ttH\ cross-section measurement in the \hyy\ decay channel are those related to the parton-shower modelling in the \ttH\ simulation, which are evaluated by comparing the shower and hadronization modelling of \textsc{Pythia8} with \textsc{Herwig7}~\cite{Bahr:2008pv,Bellm:2013hwb}, and correspond to a relative uncertainty of 8\% in the \ttH\ cross-section measurement, and the modelling uncertainty in the Higgs boson plus heavy-flavour background (4\%). The dominant experimental uncertainties are related to the reconstruction of the jet energy (5\%), the photon isolation requirements (4\%), and the photon energy resolution (6\%) and scale (4\%).
 
This analysis is about 50\% more sensitive than the one in Ref.~\cite{HIGG-2016-21} for the same integrated luminosity, with the two regions (\had\ and \lep) achieving similar sensitivity.
The improvements include new reconstruction algorithms, the relaxed requirements on jets and $b$-tagged jets, and a BDT-based instead of a cut-based selection for the \lep\ region.
The largest sensitivity improvement (about 30\%) is achieved by using four-momentum information of photons, jets and leptons, as well as $b$-tagging information of jets, as input to the BDT.
Both the \had\ BDT and the \lep\ BDT use the scaled photon $\pt$/$m_{\gamma\gamma}$ observable to prevent the diphoton mass being used as a discriminating variable by the BDT.
This is further verified using fits of the functional forms chosen in each BDT bin in several additional control regions in data and simulation, and no evidence of a bias is found.

Figure~\ref{fig:myy_ttH_weighted} shows the observed \myy\ distribution in the \ttH-sensitive BDT bins. For illustration purposes, events are weighted by $\ln(1 + S_{90}/B_{90})$, where $S_{90}$ ($B_{90}$) for each BDT bin is the expected \ttH\ signal~\cite{Raitio:1978pt,Kunszt:1984ri,Beenakker:2001rj,Beenakker:2002nc,Dawson:2002tg,Dawson:2003zu,Yu:2014cka,Frixione:2014qaa,Frixione:2015zaa,deFlorian:2016spz,powheg1,powheg2,powheg3} (background) in the smallest $m_{\gamma\gamma}$ window containing 90\% of the expected signal.
Both the signal-plus-background and background-only curves shown here are obtained from the weighted sum of the individual curves in each BDT bin.
The expected and observed event yields are presented in Table~\ref{tab:yields} and shown in Figure~\ref{fig:stacks}. In Figure~\ref{fig:stacks}, a \ttH\ signal strength $\mu = \sigma / \sigma_\mathrm{SM}$ of 1.4 is assumed. The total number of fitted \ttH\ signal events in the mass range 105~\GeV~$<m_{\gamma\gamma}$~$<$~160~\GeV\ is $36 ^{+12} _{-11}$.
For 13~\tev\ data corresponding to an integrated luminosity of 79.8~\ifb, the expected significance of the $t\bar{t}H$ signal in the \hgg\ channel is \SignGammaGammaExp\ standard deviations. The significance of the observed \ttH\ signal is \SignGammaGammaObs\ standard deviations. The expected significance in the \had\ (\lep) region is \SignGammaGammaHadExp\ (\SignGammaGammaLepExp) standard deviations, while the observed significance in the \had\ (\lep) region is \SignGammaGammaHadObs\ (\SignGammaGammaLepObs) standard deviations.
 
% End of text imported from the .//sections/gammagamma.tex input file
\section{\hfourl}
% The next lines are included from the .//sections/fourl.tex input file
\begin{table}
\renewcommand*{\arraystretch}{1.2}
 
\centering
\caption{\label{tab:yields} Observed number of events in the different bins of the \hyy\ and \hfourl\ searches, using 13~\tev\ data corresponding to an integrated luminosity of 79.8~\ifb. The observed yields are compared with the sum of expected \ttH\ signal, normalized to the SM prediction, background from non-\ttH\ Higgs boson production and other background sources,
with the systematic uncertainties assigned to the observed result in the \hyy\ analysis,
and expected  systematic uncertainties in the \hfourl\ analysis.
The numbers for \hyy\ are counted in the smallest $m_{\gamma\gamma}$ window containing 90\% of the expected signal. The numbers for \hfourl\ are derived in a four-lepton mass window of 115~\gev\ $<$ \mfourl\ $<$ 130~\gev.
In the \hyy\ analysis, the background yield is extracted from the fit with freely floating signal.
The BDT bins are in descending order of signal purity.
}
\begin{tabular}{l|S[table-format=1.3(4)]S[table-format=1.4(5)]S[table-format=1.3(4)]S[table-format=1.3(4)]|r} 
\hline\hline
& \multicolumn{4}{c|}{Expected} & \multicolumn{1}{c}{Observed} \\
Bin & \ttH\ (signal) & \multicolumn{1}{c}{Non-\ttH\ Higgs} & \multicolumn{1}{c}{Non-Higgs} & \multicolumn{1}{c|}{Total} &  \multicolumn{1}{c}{Total} \\
\hline
\multicolumn{6}{c}{\hyy}\\
\hline
\had\ 1 & 4.2 \pm 1.1     & 0.49 \pm 0.33 & 1.8 \pm 0.5 & 6.4 \pm 1.3  & 10 \\
\had\ 2 & 3.4 \pm 0.7   & 0.7 \pm 0.6 & 7.5 \pm 1.1   & 11.6 \pm 1.5 & 14 \\
\had\ 3 & 4.7 \pm 0.9   & 2.0 \pm 1.7   & 32.9 \pm 2.2  & 39.6 \pm 3.2  & 47 \\
\had\ 4 & 3.0 \pm 0.5   & 3.2 \pm 3.1   & 55.0 \pm 2.8  & 61 \pm 5  & 67 \\
\lep\ 1 & 4.5 \pm 1.0     & 0.24 \pm 0.09 & 2.2 \pm 0.6 & 6.9 \pm 1.2   & 7  \\
\lep\ 2 & 2.2 \pm 0.4   & 0.27 \pm 0.10 & 4.6 \pm 0.9 & 7.1 \pm 1.0   & 7  \\
\lep\ 3 & 0.82 \pm 0.18   & 0.30 \pm 0.13 & 4.6 \pm 0.9 & 5.7 \pm 0.9 & 5  \\
\hline
\multicolumn{6}{c}{\hfourl}\\
\hline
\had\ 1 & 0.169 \pm 0.031  &  0.021 \pm 0.007  & 0.008 \pm 0.008  &  0.198 \pm 0.033  & 0 \\
\had\ 2 & 0.216 \pm 0.032  &  0.20 \pm  0.09  &  0.22 \pm  0.12  &   0.63 \pm  0.16  & 0 \\
\lep\   & 0.212 \pm 0.031  &  0.0256 \pm 0.0023  &  0.015 \pm 0.013 &  0.253 \pm 0.034  & 0 \\
\hline\hline
\end{tabular}
\end{table}
In the \hfourl\ analysis, using the same data as in the \hyy\ analysis, events with at least four isolated leptons (four electrons, four muons, or two electrons and two muons) corresponding to two same-flavour opposite-charge pairs are selected. The four-lepton invariant mass is required to be in a window of 115~\gev\ $<$ \mfourl\ $<$ 130~\gev. To search for \ttH\ events, at least one jet is required, with \pt $>$ 30~\gev\ and containing a $b$-hadron identified using a $b$-tagging algorithm with an efficiency of 70\%. The event selection is described in more detail in Ref.~\cite{Aaboud:2017vzb}.
The current analysis improves the expected \ttH\ significance by defining two signal regions, and by applying a BDT in one of them. A `\had' region enriched in hadronic top-quark decays is formed by requiring at least three additional jets and zero additional isolated leptons, and a `\lep' region enriched in semileptonic top-quark decays is formed by requiring at least one additional jet and at least one additional isolated lepton. The main backgrounds in both regions are \ttW, \ttZ, and non-\ttH\ Higgs boson production (ggF and $tH$ for the \had\ and $tH$ for the \lep\ region), estimated from simulation. The same event generators and cross sections are used as in the \hyy\ analysis.
Uncertainties due to parton distribution functions (PDF) and $\alphas$, and missing higher-order corrections are considered.
To account for the theoretical uncertainty in the radiation of additional heavy-flavour jets, a 100\% uncertainty is assigned to the predicted ggF yields.
In the \had\ region, a BDT~\cite{Hocker:2007ht} is employed to separate the \ttH\ signal from the background. Eleven observables are used, including the invariant mass, the dijet \pt, and the difference in pseudorapidity $\Delta \eta$ of the two leading jets, as well as the difference between the $\eta$ of the four-lepton system and the average $\eta$ of the two leading jets.
Further input observables are \MET, the angular separation $\Delta R$ between the four-lepton system and the leading jet, as well as between the dilepton pair with invariant mass closest to the $Z$ boson mass and the leading jet, the scalar sum of the \pt\ of the jets in the event, the number of jets, the number of $b$-tagged jets, and the value of the leading-order matrix element describing the Higgs boson decay~\cite{Aaboud:2017vzb}. This matrix-element value will be larger for the leptons from the Higgs boson decay than for those from the \ttZ\ and \ttW\ background.
The output discriminant of this BDT is divided into two bins, which are chosen to maximize the expected \ttH\ significance in the \had\ region. The bin with the higher values of the BDT discriminant and the \lep\ region are expected to have a \ttH\ signal purity of more than 80\%. The other BDT bin is expected to have a \ttH\ signal purity of about 35\%.
 
The observed events and expected background yields in the two \had\ BDT bins and the \lep\ region, in a four-lepton invariant mass window of 115~\gev~$<$~\mfourl~$<$~130~\gev, are used as input to a likelihood fit that extracts the \ttH\ yield.
The expected dominant uncertainties in the cross section are due to the parton-shower modelling affecting the acceptance of the selection, and to the cross-section uncertainty in the Higgs boson plus heavy-flavour background (about 10\% each).
The leading experimental uncertainty arises from the calibration of the jet energy scale (6\%).
The expected and observed numbers of events are presented in Table~\ref{tab:yields}. No event is observed.
The expected significance is \SignFourLExp\ standard deviations.

% End of text imported from the .//sections/fourl.tex input file
\section{Combination}
% The next lines are included from the .//sections/combination.tex input file
The \ttH\ searches in the \hyy\ and \hfourl\ decay channels are combined with the \hbb\ and multilepton searches from Refs.~\cite{HIGG-2017-03,HIGG-2017-02}. These analyses use a dataset corresponding to an integrated luminosity of 36.1~\ifb\ at $\sqrt{s}=13$~\tev, and find observed (expected) significances of \SignbbObs\ (\SignbbExp) standard deviations for \hbb\ and \SignMLObs\ (\SignMLExp) for the multilepton search.
The combination is performed using the profile likelihood method described in Ref.~\cite{ATL-PHYS-PUB-2011-011-fixed}, based on simultaneous fits to the signal regions and control regions of the individual analyses.
The overlap between the selected events in the different analyses is found to be negligible. The asymptotic approximation used in the fit is verified with pseudo-experiments, and the results are corrected if necessary.
The effect of systematic uncertainties in the predicted yields and distributions is incorporated into the statistical model through nuisance parameters. The correlation scheme of all systematic uncertainties between the \hbb\ and multilepton analyses, as well as the correlation scheme of the theory uncertainties between all channels are the same as in Ref.~\cite{HIGG-2017-02}. Since the \hyy\ and \hfourl\ analyses employ improved reconstruction software compared with the \hbb\ and multilepton analyses, the correlations between the experimental systematic uncertainties are evaluated for each source individually. Some components of the systematic uncertainties in the luminosity, the jet energy scale, the electron/photon resolution and energy scale, and in the electron reconstruction and identification efficiencies are correlated between the channels.
All Higgs boson production processes other than \ttH, including Higgs boson production in association with a single top quark,
are considered as background and their cross sections are fixed to the SM predictions~\cite{deFlorian:2016spz}.
The respective cross-section uncertainties are considered as systematic uncertainties.
The total \ttH\ cross section is extracted assuming SM branching fractions and using the detector acceptance and efficiencies predicted from the \ttH\ simulation discussed above. The respective uncertainties are included in the fit.

A combination is also performed  with the \ttH\ searches based on datasets corresponding to integrated luminosities of 4.5~\ifb\ at $\sqrt{s}=7$~\TeV\ and 20.3~\ifb\ at $\sqrt{s}=8$~\tev~\cite{HIGG-2014-06}.  The combined observable is the signal strength $\mu = \sigma / \sigma_{\mathrm{SM}}$. The SM cross-section expectations $\sigma_{\mathrm{SM}}$ and branching ratios used in the 7 and 8~\tev\ analyses are updated with the values in Ref.~\cite{deFlorian:2016spz}, while their uncertainties are not changed. Theoretical uncertainties in the SM cross-section prediction for \ttH\ are included in the signal-strength extraction. The branching-fraction uncertainties and the uncertainties due to missing higher-order corrections in the \ttH\ cross-section prediction are correlated between the 7 and 8~\tev\ and 13~\tev\ analyses.  Furthermore, the relevant uncertainties in the electron/photon energy scale and resolution are correlated.

% End of text imported from the .//sections/combination.tex input file
\section{Results}
% The next lines are included from the .//sections/resultsconclusions.tex input file
Table~\ref{tab:sys} shows a summary of the systematic uncertainties in the 13~\tev\ \ttH\ production cross-section measurement. The dominant uncertainties arise from the modelling of the \ttbar + heavy-flavour processes in the \hbb\ analysis~\cite{HIGG-2017-03} and the modelling of the \ttH\ process, which affects the acceptance of the selection in all analyses. Further important uncertainties come from uncertainties in the estimate of leptons from heavy-flavour decays, conversions or misidentified hadronic jets, mainly in the multilepton analysis~\cite{HIGG-2017-02}, and in the jet energy scale and resolution in all analyses. The jet, electron, and photon uncertainties, as well as the uncertainties associated with hadronically decaying $\tau$-leptons, include uncertainties in the reconstruction and identification efficiencies, as well as in the energy scale and resolution. The $\tau$-lepton uncertainty affects the multilepton analysis. The Monte Carlo (MC) statistical uncertainty is due to limited numbers of simulated events in the \hbb\ and multilepton analyses.

\begin{table}
\centering
\caption{\label{tab:sys} Summary of the systematic uncertainties affecting the combined \ttH\ cross-section measurement at 13~\tev. Only systematic uncertainty sources with at least 1\% impact are listed. The fake-lepton uncertainty is due to the estimate of leptons from heavy-flavour decay, conversions or misidentified hadronic jets. The jet, electron, and photon uncertainties, as well as the uncertainties associated with hadronically decaying $\tau$-leptons, include those in reconstruction and identification efficiencies, as well as in the energy scale and resolution. The Monte Carlo (MC) statistical uncertainty is due to limited numbers of simulated events.
More detailed descriptions of the sources of the systematic uncertainties  are given in Refs.~\cite{HIGG-2017-03,HIGG-2017-02}.
}
\begin{tabular}{lr}
\hline\hline
Uncertainty source   & $\Delta\sigma_{\ttH}/\sigma_{\ttH}$ [\%] \\
\hline
Theory uncertainties (modelling)                             & 11.9 \\
~~~~~\ttbar\ + heavy flavour                                    & 9.9 \\
~~~~~\ttH\                                         & 6.0 \\
~~~~~Non-\ttH\ Higgs boson production                & 1.5 \\
~~~~~Other background processes                  & 2.2 \\
\hline
Experimental uncertainties  & 9.3 \\
~~~~~Fake leptons                                           & 5.2 \\
~~~~~Jets, \MET                  & 4.9 \\
~~~~~Electrons, photons                              & 3.2\\
~~~~~Luminosity                                             & 3.0 \\
~~~~~$\tau$-leptons                                           & 2.5 \\
~~~~~Flavour tagging                                        & 1.8 \\
\hline
MC statistical uncertainties                                    & 4.4 \\
\hline\hline
\end{tabular}
\end{table}
\begin{figure}
\centering
\includegraphics[width=0.55\textwidth]{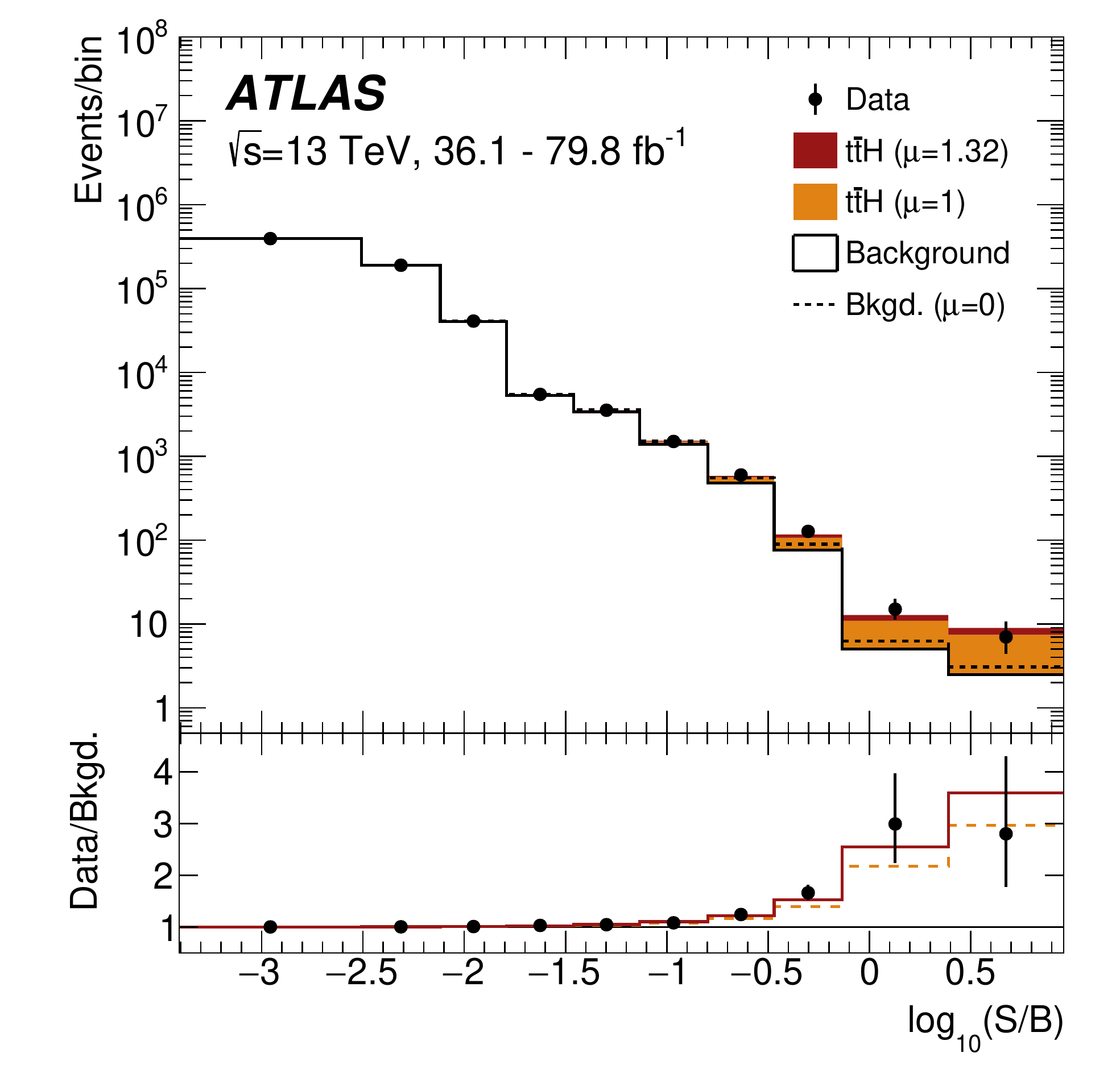}
\caption{Observed event yields in all analysis categories in up to 79.8~\ifb\ of 13~\TeV\ data. The background yields correspond to the observed fit results, and the signal yields are shown for both the observed results ($\mu = 1.32$) and the SM prediction ($\mu = 1$). The discriminant bins in all categories are ranked by $\log_{10}(S/B)$, where $S$ is the signal yield and $B$ the background yield extracted from the fit with freely floating signal, and combined such that $\log_{10}(S+B)$ decreases approximately linearly. For the \hyy\ analysis, only events in the smallest $m_{\gamma\gamma}$ window containing 90\% of the expected signal are considered.
The lower panel shows the ratio of the data to the background estimated from the fit with freely floating signal, compared to the expected distribution including the signal assuming $\mu = 1.32$ (full red) and $\mu = 1$ (dashed orange). The error bars on the data are statistical.
\label{fig:log_sb}}
\end{figure}
\begin{figure}[]
\centering
\includegraphics[width=0.7\linewidth]{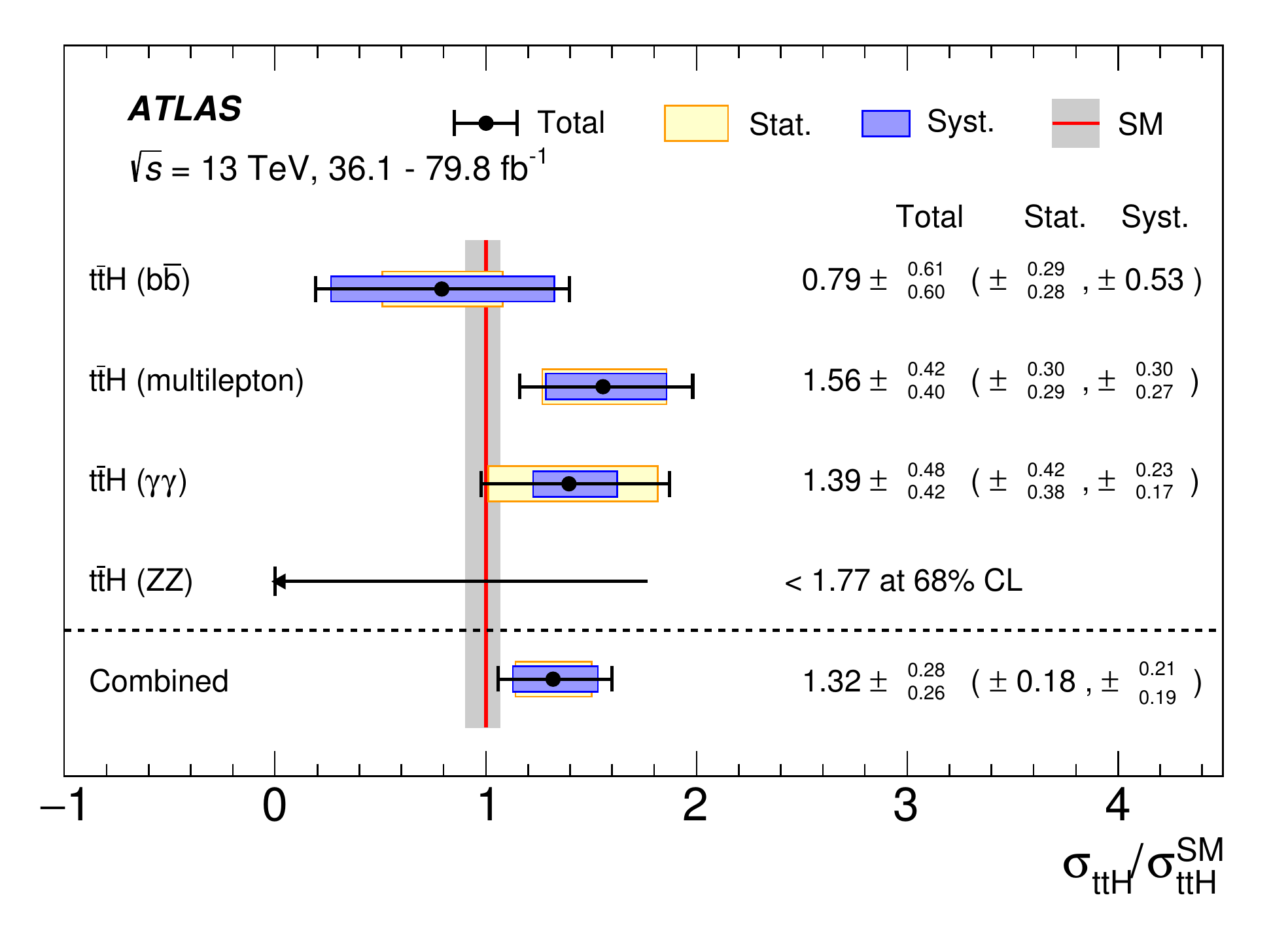}
\caption{Combined \ttH\ production cross section, as well as cross sections measured in the individual analyses, divided by the SM prediction. The \yy\ and \ZZfourl\ analyses use 13~\tev\ data corresponding to an integrated luminosity of 79.8~\ifb, and the multilepton and \bb\ analyses use data corresponding to an integrated luminosity of 36.1~\ifb. The black lines show the total uncertainties, and the bands indicate the statistical and systematic uncertainties.
The red vertical line indicates the SM cross-section prediction, and the grey band represents the PDF+$\alphas$ uncertainties and the uncertainties due to missing higher-order corrections.
\label{fig:results}}
\end{figure}
\begin{figure}[]
\centering
\includegraphics[width=0.65\linewidth]{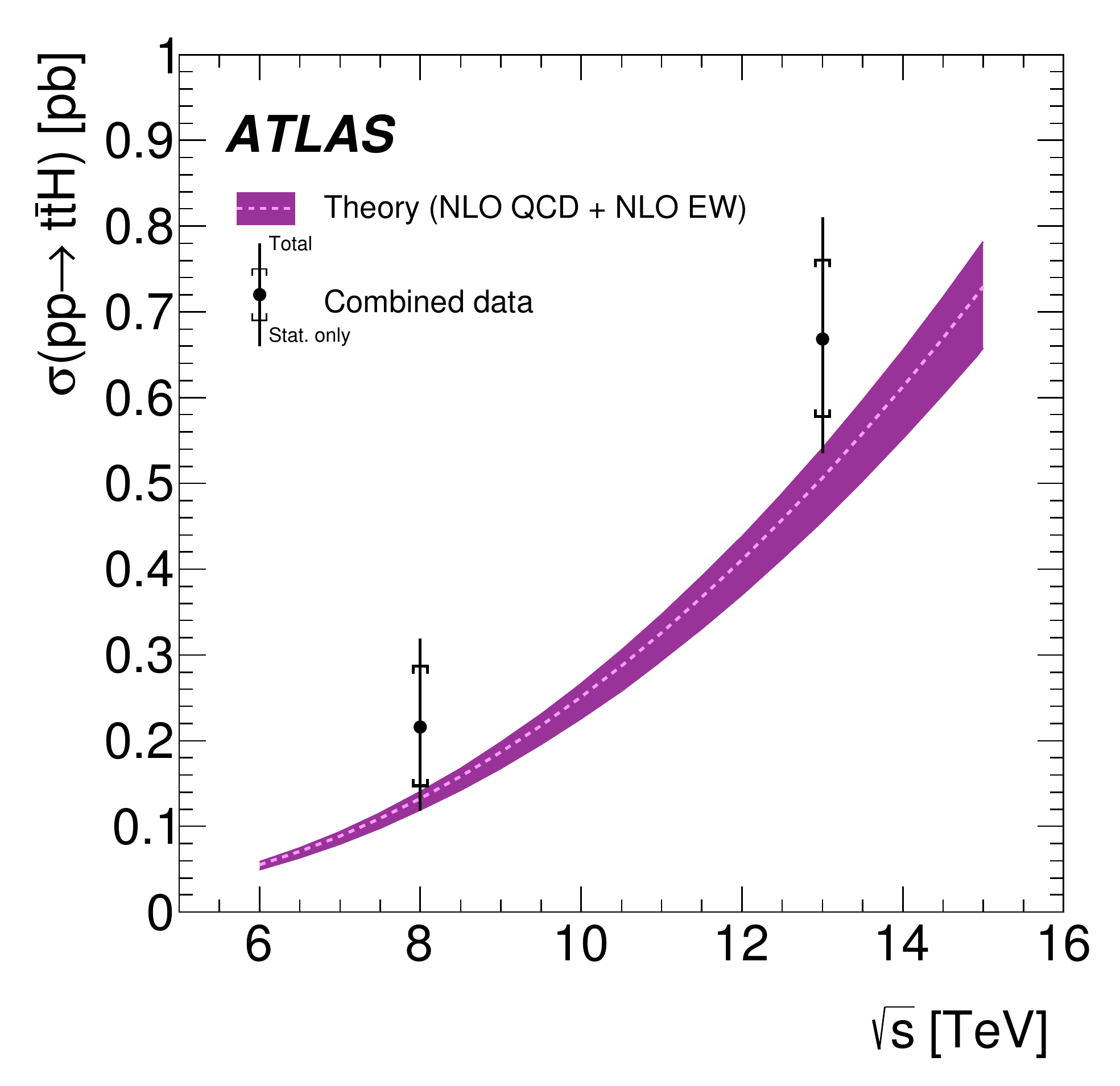}
\caption{Measured \ttH\ cross sections in $pp$ collisions at centre-of-mass energies of  8~\TeV\ and 13~\tev. Both the total and statistical-only uncertainties are shown. The measurements are compared with the SM prediction.
The band around the prediction represents the PDF+$\alphas$ uncertainties and the uncertainties due to missing higher-order corrections.
\label{fig:sqrt}}
\end{figure}
\begin{table}
\renewcommand*{\arraystretch}{1.2}
\centering
\caption{\label{tab:results} Measured total \ttH\ production cross sections at 13~\tev, as well as observed (Obs.) and expected (Exp.) significances (sign.) relative to the background-only hypothesis. The results of the individual analyses, as well as the combined results are shown.
Since no event is observed in the \hfourl\ decay channel, an observed upper limit is set at 68\% confidence level on the \ttH\ production cross section in that channel using pseudo-experiments.    }
\begin{tabular}{lcccc}
\hline\hline
Analysis & Integrated &  \ttH\ cross & Obs. & Exp.  \\
& luminosity [\ifb]  & section [fb] & sign. & sign.  \\
\hline
\hyy\  & 79.8 & \XSGammaGamma & \SignGammaGammaObs $\sigma$ & \SignGammaGammaExp $\sigma$ \\
$H \rightarrow$~multilepton & 36.1 & \XSML & \SignMLObs $\sigma$& \SignMLExp $\sigma$\\
\hbb\  & 36.1 & \XSbb & \SignbbObs $\sigma$& \SignbbExp $\sigma$\\
\hfourl & 79.8 & $<$\XSFourL\ (68\% CL) & \SignFourLObs $\sigma$& \SignFourLExp $\sigma$\\
\hline
Combined (13~\tev) & 36.1$-$79.8 & \XSComb & \SignCombRunTwoObs $\sigma$& \SignCombRunTwoExp $\sigma$\\
Combined (7, 8, 13~\tev) & 4.5, 20.3, 36.1$-$79.8   & $-$ & \SignCombRunOneRunTwoObs $\sigma$& \SignCombRunOneRunTwoExp $\sigma$\\
\hline\hline
\end{tabular}
\end{table}
Using 13~\tev\ data, the likelihood fit to extract the \ttH\ signal yield in the \hyy, \hfourl, \hbb, and multilepton analyses results in an observed (expected) excess relative to the background-only hypothesis of \SignCombRunTwoObs\ (\SignCombRunTwoExp) standard deviations. A combined fit using the  7, 8, and 13~\tev\ analyses  gives an observed (expected) significance of \SignCombRunOneRunTwoObs\ (\SignCombRunOneRunTwoExp) standard deviations. Table~\ref{tab:results} shows the significances of the individual and combined analyses relative to the background-only hypothesis. Figure~\ref{fig:log_sb}  shows the combined event yields in all analysis categories as a function of $\log_{10}(S/B)$, where $S$ is the expected signal yield and $B$ the background yield extracted from the fit with freely floating signal. A clear \ttH\ signal-like excess over the background is visible for high $\log_{10}(S/B)$.

Based on the analyses performed at 13~\tev, the measured total cross section for \ttH\ production is \XSComb~fb, in agreement with the SM prediction of 507$^{+35}_{-50}$~fb~\cite{Raitio:1978pt,Kunszt:1984ri,Beenakker:2001rj,Beenakker:2002nc,Dawson:2002tg,Dawson:2003zu,Yu:2014cka,Frixione:2014qaa,Frixione:2015zaa,deFlorian:2016spz}, which is calculated to next-to-leading-order accuracy (both QCD and electroweak).
The cross section extracted in the combined likelihood fit, as well as the results from the individual analyses, are shown in Table~\ref{tab:results},
while their ratios to the SM predictions are displayed in Figure~\ref{fig:results}.
The measured total cross section for \ttH\ production at 8~\tev\ is \XSCombEight~fb.
Figure~\ref{fig:sqrt} shows the \ttH\ production cross sections measured in $pp$ collisions at centre-of-mass energies of 8 and 13~\tev, compared to the SM predictions.
 
\FloatBarrier
\section{Conclusion}
Using proton--proton collision data at centre-of-mass energies of 7, 8, and 13~\tev, produced by the Large Hadron Collider and recorded with the ATLAS detector,
the production of the Higgs boson in association with a top quark pair is observed with a significance of \SignCombRunOneRunTwoObs\ standard deviations relative to the background-only hypothesis. The expected significance is \SignCombRunOneRunTwoExp\ standard deviations.
The \ttH\ production cross section at 13~\tev\ is measured in data corresponding to integrated luminosities of up to 79.8~\ifb\ to be \XSComb~fb, in agreement with the Standard Model prediction.
This constitutes a direct observation of the Yukawa coupling between the Higgs boson and the top quark.

% End of text imported from the .//sections/resultsconclusions.tex input file
 
\section*{Acknowledgements}
 
% The next lines are included from the .//acknowledgements/Acknowledgements.tex input file
 
We thank CERN for the very successful operation of the LHC, as well as the
support staff from our institutions without whom ATLAS could not be
operated efficiently.
 
We acknowledge the support of ANPCyT, Argentina; YerPhI, Armenia; ARC, Australia; BMWFW and FWF, Austria; ANAS, Azerbaijan; SSTC, Belarus; CNPq and FAPESP, Brazil; NSERC, NRC and CFI, Canada; CERN; CONICYT, Chile; CAS, MOST and NSFC, China; COLCIENCIAS, Colombia; MSMT CR, MPO CR and VSC CR, Czech Republic; DNRF and DNSRC, Denmark; IN2P3-CNRS, CEA-DRF/IRFU, France; SRNSFG, Georgia; BMBF, HGF, and MPG, Germany; GSRT, Greece; RGC, Hong Kong SAR, China; ISF, I-CORE and Benoziyo Center, Israel; INFN, Italy; MEXT and JSPS, Japan; CNRST, Morocco; NWO, Netherlands; RCN, Norway; MNiSW and NCN, Poland; FCT, Portugal; MNE/IFA, Romania; MES of Russia and NRC KI, Russian Federation; JINR; MESTD, Serbia; MSSR, Slovakia; ARRS and MIZ\v{S}, Slovenia; DST/NRF, South Africa; MINECO, Spain; SRC and Wallenberg Foundation, Sweden; SERI, SNSF and Cantons of Bern and Geneva, Switzerland; MOST, Taiwan; TAEK, Turkey; STFC, United Kingdom; DOE and NSF, United States of America. In addition, individual groups and members have received support from BCKDF, the Canada Council, CANARIE, CRC, Compute Canada, FQRNT, and the Ontario Innovation Trust, Canada; EPLANET, ERC, ERDF, FP7, Horizon 2020 and Marie Sk{\l}odowska-Curie Actions, European Union; Investissements d'Avenir Labex and Idex, ANR, R{\'e}gion Auvergne and Fondation Partager le Savoir, France; DFG and AvH Foundation, Germany; Herakleitos, Thales and Aristeia programmes co-financed by EU-ESF and the Greek NSRF; BSF, GIF and Minerva, Israel; BRF, Norway; CERCA Programme Generalitat de Catalunya, Generalitat Valenciana, Spain; the Royal Society and Leverhulme Trust, United Kingdom.
 
The crucial computing support from all WLCG partners is acknowledged gratefully, in particular from CERN, the ATLAS Tier-1 facilities at TRIUMF (Canada), NDGF (Denmark, Norway, Sweden), CC-IN2P3 (France), KIT/GridKA (Germany), INFN-CNAF (Italy), NL-T1 (Netherlands), PIC (Spain), ASGC (Taiwan), RAL (UK) and BNL (USA), the Tier-2 facilities worldwide and large non-WLCG resource providers. Major contributors of computing resources are listed in Ref.~\cite{ATL-GEN-PUB-2016-002}.
 
% End of text imported from the .//acknowledgements/Acknowledgements.tex input file

\printbibliography

\clearpage
\input{atlas_authlist} 
\end{document}

%% file: atlas_authlist.tex
% ATLAS Collaboration author list
% Reference date of HIGG-2018-13 is 2018-05-14
% Author list last updated on date 13-AUG-18
% Data extracted on 13-Aug-2018 for paper reference HIGG-2018-13
% at 11:25pm
 
\begin{flushleft}
{\Large The ATLAS Collaboration}

\bigskip

M.~Aaboud$^\textrm{\scriptsize 34d}$,    
G.~Aad$^\textrm{\scriptsize 99}$,    
B.~Abbott$^\textrm{\scriptsize 124}$,    
O.~Abdinov$^\textrm{\scriptsize 13,*}$,    
B.~Abeloos$^\textrm{\scriptsize 128}$,    
D.K.~Abhayasinghe$^\textrm{\scriptsize 91}$,    
S.H.~Abidi$^\textrm{\scriptsize 164}$,    
O.S.~AbouZeid$^\textrm{\scriptsize 39}$,    
N.L.~Abraham$^\textrm{\scriptsize 153}$,    
H.~Abramowicz$^\textrm{\scriptsize 158}$
% ...
\bigskip
\\

% ...
$^{13}$Institute of Physics, Azerbaijan Academy of Sciences, Baku; Azerbaijan.\\
% ...
$^{34}$$^{(a)}$Facult\'e des Sciences Ain Chock, R\'eseau Universitaire de Physique des Hautes Energies - Universit\'e Hassan II, Casablanca;$^{(b)}$Centre National de l'Energie des Sciences Techniques Nucleaires (CNESTEN), Rabat;$^{(c)}$Facult\'e des Sciences Semlalia, Universit\'e Cadi Ayyad, LPHEA-Marrakech;$^{(d)}$Facult\'e des Sciences, Universit\'e Mohamed Premier and LPTPM, Oujda;$^{(e)}$Facult\'e des sciences, Universit\'e Mohammed V, Rabat; Morocco.\\
% ...
$^{39}$Niels Bohr Institute, University of Copenhagen, Copenhagen; Denmark.\\
% ...
$^{91}$Department of Physics, Royal Holloway University of London, Egham; United Kingdom.\\
% ...
$^{99}$CPPM, Aix-Marseille Universit\'e, CNRS/IN2P3, Marseille; France.\\
% ...
$^{124}$Homer L. Dodge Department of Physics and Astronomy, University of Oklahoma, Norman OK; United States of America.\\
% ...
$^{128}$LAL, Universit\'e Paris-Sud, CNRS/IN2P3, Universit\'e Paris-Saclay, Orsay; France.\\
% ...
$^{153}$Department of Physics and Astronomy, University of Sussex, Brighton; United Kingdom.\\
% ...
$^{158}$Raymond and Beverly Sackler School of Physics and Astronomy, Tel Aviv University, Tel Aviv; Israel.\\
% ...
$^{164}$Department of Physics, University of Toronto, Toronto ON; Canada.\\
% ...
$^{a}$ Also at Borough of Manhattan Community College, City University of New York, NY; United States of America.\\
$^{b}$ Also at Centre for High Performance Computing, CSIR Campus, Rosebank, Cape Town; South Africa.\\
$^{c}$ Also at CERN, Geneva; Switzerland.\\
$^{d}$ Also at CPPM, Aix-Marseille Universit\'e, CNRS/IN2P3, Marseille; France.\\
$^{e}$ Also at D\'epartement de Physique Nucl\'eaire et Corpusculaire, Universit\'e de Gen\`eve, Gen\`eve; Switzerland.\\
% ...
$^{*}$ Deceased

\end{flushleft}

% Created with Glance <Atlas.Glance@cern.ch>